\documentclass[preprint,showpacs,preprintnumbers]{revtex4}
\usepackage{graphicx}
\usepackage{dcolumn}
\usepackage{bm}
\usepackage{amsmath}
\usepackage{amsfonts}
\usepackage{rotating}
\usepackage{xspace}

\newcommand{\cm}{\ensuremath{\rm cm^{-1}}\xspace}
\newcommand{\Eh}{\ensuremath{E_{\mathrm{h}}}\xspace}

\newcommand{\Htp}{\ensuremath{\rm H_2^+}\xspace}

\newcommand{\Http}{\ensuremath{\rm H_3^+}\xspace}

\begin{document}

\title{Vibrational states of the triplet electronic state of H$_3^+$.\\ The role of  non-adiabatic coupling and  geometrical phase.}
\author{Alexander Alijah$^{1}$ and Viatcheslav Kokoouline$^{2}$}
\affiliation{$^{1}$ Groupe de Spectrometrie Mol\'eculaire et Atmospherique, University of Reims Champagne-Ardenne, F-51687, Reims Cedex 2, France \\
$^{2}$Department of Physics, University of Central Florida, Orlando, Florida 32816, USA }

\date{\today}


\begin{abstract} 
Vibrational energies and wave functions  of the triplet state of the \Http ion have been determined. In the calculations, the ground and first excited triplet electronic states are included as well as the non-Born-Oppenheimer coupling between them. A diabatization procedure transforming the two adiabatic {\it ab initio} potential energy surfaces of the  triplet-\Http state into a $2\times2$ matrix is employed. The diabatization takes into account the non-Born-Oppenheimer coupling and the effect of the geometrical phase due to the conical intersection between the two adiabatic potential surfaces.  The results are compared to the calculation involving only the lowest adiabatic potential energy surface of  the triplet-\Http ion and neglecting the geometrical phase. The energy difference between results with and without the non-adiabatic coupling and the geometrical phase is about a wave number for the lowest vibrational levels.
\end{abstract}


\maketitle

\section{Introduction}

Hydrogen is the most abundant element in the universe. The most important hydrogenic compounds are the dihyrogen molecule
and its protonated form, \Http. The latter is a strong acid and very reactive, but does exist in large amounts in low density 
environments such as the interstellar space. The role it plays in interstellar chemistry has been reviewed by Oka~\cite{oka06b,OKA13:8738}.
The ground electronic state of \Http is a singlet, $\tilde{X} \, ^1A_1$, with the minimum of the potential energy surface (PES) at the equilateral 
geometry (EG) of three nuclei with the internuclear distance $r=1.6504$ bohr ($a_0$ below). 

Above the ground singlet state of \Http, there is a metastable triplet state, $\tilde{a} \, ^3\Sigma_u^+$.
It has never been observed but has been predicted to exist in 1974 by Schaad and Hicks~\cite{SCH74:1934}
and explored theoretically. The first complete potential energy surfaces (PES) and vibrational and rovibrational calculations
were published only 27 years later and independently by two groups~\cite{FRI01:1183,SAN01:2182}.
The lowest triplet PES is characterized by three minima at symmetric linear configurations with $r_1=r_2=r_3/2=2.4537\,a_0$.
They are separated by barriers of 2596 \cm, with configurations $r_1=1.99 \, a_0, r_2=r_3=5.336 \, a_0$
corresponding to T-shaped complexes of a $\rm H_2^+$ ion and a distant hydrogen atom. The three dissociation channels to  $\rm H_2^+ + H$
are at 2946.8 \cm. At EG, the state is degenerate with the second triplet state for symmetry reasons.
The curve of degeneracy has a minimum at $E= 14934\ \cm$ above the dissociation energy or $E= 17880.8\ \cm$ above the minimum,
with internuclar distance  $r_1=r_2=r_3=3.610 \, a_0$. 
As the molecule departs from the equilateral configuration, the degeneracy is lifted according
to the Jahn-Teller theorem~\cite{JAH37:220}. Therefore, the two triplet electronic states 
could be viewed as two components of the $^3E'$ electronic state in the $D_{3h}(M)$ molecular symmetry group,
which intersect conically at equilateral configurations.  For the simplicity of discussion, we will 
be referring the two lowest electronic triplet states as the lower ($l$) and upper ($u$) states. 
Individual fits of either of the two states and corresponding rovibrational calculations have been published~\cite{CER03:2637,ALI03:163,VIE04:253}.
A two-surface potential has also been reported~\cite{varandas05}. 
For an overview on triplet \Http see Alijah and Varandas~\cite{ALI06:2889}.

Vibrational energies obtained taking into account only the lowest triplet state (one-state calculations) and neglecting the interaction 
with the upper state may not be precise even though the upper state is located at high energy, in the continuum of the lower state. 
In this study we discuss the effect of the coupling with the upper state on the vibrational energies 
of the lower state and in particular the effect of the geometrical phase.

\section{Theoretical approach }
\label{sec:theory}

\subsection{Diabatization representation of the potential}

\begin{figure}
\includegraphics[width=14cm]{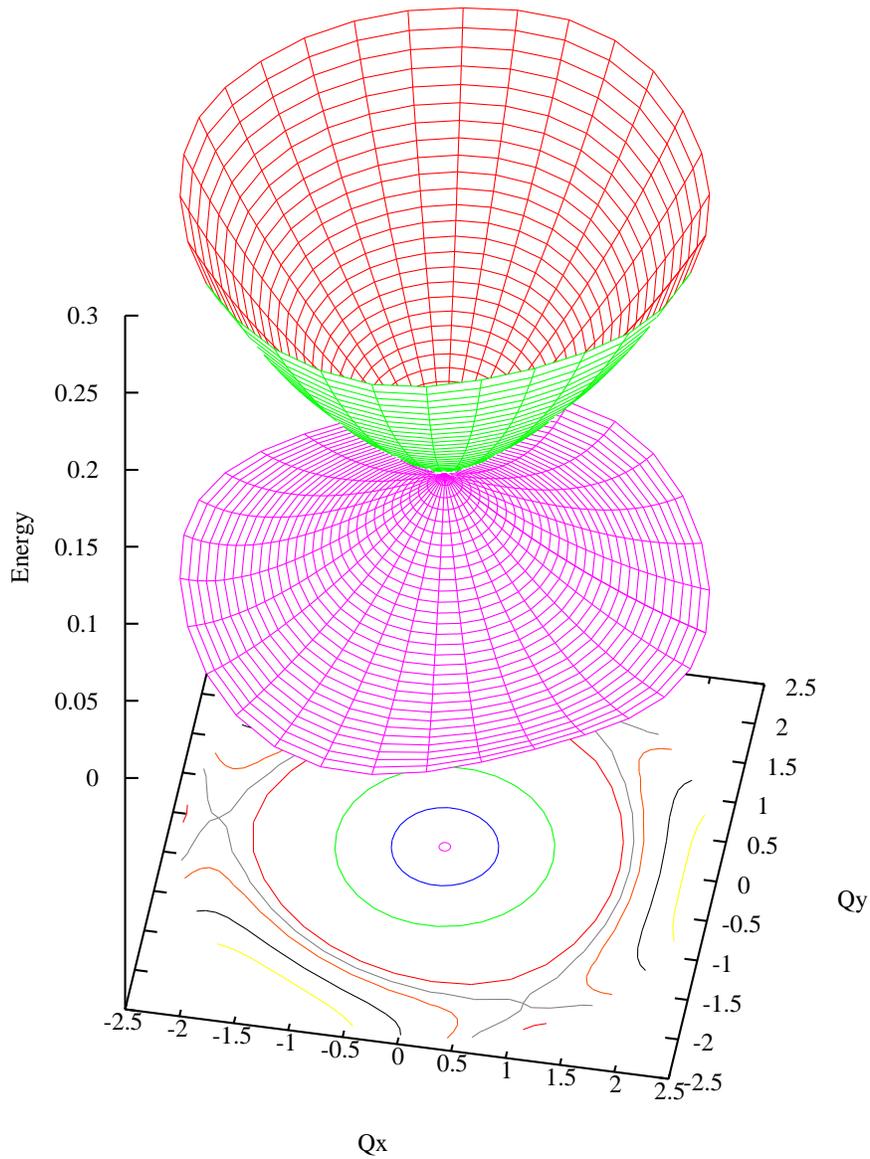}
\vspace{-1.5cm}
\caption{Born-Oppenheimer (adiabatic) potential energy surfaces of the two components of the \Http triplet state shown as a function of two asymmetric stretch coordinates, $Q_x$ and $Q_y$ of Eq.~(\ref{eq:symm_coor}) for a fixed value of the symmetric stretch coordinate $Q_1=5.3$ $a_0$. The three minima of the shown part of the lowest PES are situated at linear isosceles geometries of the three nuclei. }
\label{fig:H3t_pot}
\end{figure}

The triplet-\Http PES calculated  by Varandas {\em et al.}~\cite{varandas05} is utilized in
the present work. It provides an accurate representation of the two components, in which their 
degeneracy along the line of the conical intersection and the linear splitting 
in energy with the symmetry breaking vibrational coordinate near the intersection seam 
is guaranteed by construction. This PES is shown in Fig.~\ref{fig:H3t_pot}.
In order to take into account the non-adiabatic interaction between the two components, we employ an approach similar to the one discussed in Ref.~\cite{blandon09}. Namely, we transform the two-component adiabatic (Born-Oppenheimer approximation) PES of Ref. \cite{varandas05} into a diabatic form, where the potential is written in the basis of two degenerate electronic functions $\vert +\rangle$ and $\vert -\rangle$.  
For the diabatic electronic basis, we use the electronic states evaluated at the EG \cite{longuet61}. At the EG, the electronic functions of the  triplet-\Http ion are doubly degenerate so we can choose any linear combinations of the two functions as the basis. Following Longuet-Higgins~\cite{longuet61}, we choose the linear combinations that are orthogonal to each other and transform in the $C_{3v}(M)$ symmetry group as $E_+$ and $E_-$ components of the $E$ representation:
\begin{eqnarray}
\label{eq:Eplus_minus}
(123)E_+=\omega E_+\,;\ (12)E_+=E_- \,;\nonumber\\
(123)E_-=\omega^* E_-\,;\ (12)E_-=E_+ \,,
\end{eqnarray}
where $(123)$ and $(12)$ are the cyclic and binary permutations of \Http nuclei \cite{bunker98} and $\omega=e^{2\pi i/3}$. 
In this chapter, we use the symmetry labels of the molecular symmetry group $C_{3v}(M)$, since the odd-parity representations
of $D_{3h}(M)$ are not needed.

In the diabatic basis, the potential of interaction between the three nuclei is  written as a $2\times2$ matrix with matrix elements dependent on geometry.  Because the energy of two components of the potential is the same at EG, the diagonal elements of the diabatic potential are the same. In a general form the potential is written as
\begin{eqnarray}
\label{eq:couple_pots}
\hat V = \left ( \begin{array}{cc}
A & C e^{if} \\ 
C e^{-if} & A \end{array}\right ),
\end{eqnarray}
where $A$, $C$, and $f$ are real-valued functions of internal coordinates.  The functions $A$ and $C$ are transformed in the $C_{3v}(M)$ molecular symmetry group  according to the $A_1$ irreducible representation. The phase $f$ of the coupling has the following property under the $\hat C_3$ symmetry operator of the $C_{3v}(M)$ group: $\hat C_3 f=f+2\pi/3$ , i.e. the non-diagonal elements transform as $E_+$ and $E_-$. The $A$ and $C$ functions are uniquely determined from the two Born-Oppenheimer components, $V_l$ and $V_u$, of the triplet state PES: $A=(V_l+V_u)/2$ and $C=(V_l-V_u)/2$. The actual form of the function $f$ can be derived near the EG using the diabatization procedure suggested by Longuet-Higgins \cite{longuet61}. In the diabatization procedure, the phase of the non-diagonal coupling is equal to the phase $\alpha$ of  asymmetric distortion \cite{longuet61,kokoouline03b} in symmetry-adapted coordinates. 
The symmetry-adapted coordinates $Q_1,Q_x$ and $Q_y$ are related to three internuclear distances $r_1,r_2,r_3$ as
\begin{eqnarray}
\label{eq:symm_coor}
Q_1=(r_1+r_2+r_3)/\sqrt{3}\,,\nonumber\\
Q_x=(2r_3-r_1-r_2)/\sqrt{3}\,,\\
Q_y=r_2-r_1\nonumber\,.
\end{eqnarray}
$Q_x$ and $Q_y$ become degenerate at $D_{3h}$ configurations and may be expressed more conveniently in polar form,
\begin{eqnarray}
\label{eq:symm_coor2}
\rho=\sqrt{Q_x^2+Q_y^2}\,,\nonumber\\
Q_x=\rho\cos\alpha\,,\\
Q_y=\rho\sin\alpha\nonumber\,.
\end{eqnarray}
As suggested by Longuet-Higgins, we set $f=\alpha$. Although this form of $f$ is derived near the EG, we will also use it in the region far from the EG. It is a good approximation because (1) the transition between adiabatic states occur only near the EG, where energies of the two PES components approaches, and (2) far from the EG, the phase factor $e^{\pm if}$ does not play a significant role as long as the symmetry of $f$ in the group $C_{3v}(M)$, mentioned above, is satisfied.

Notice that the basis defined above is diabatic with respect to the asymmetric stretch coordinates only but  it still depends on the symmetric stretch, i.e it is adiabatic along the coordinate.  However, for the simplicity of discussion, we refer this basis as diabatic, even if it is only partially diabatic.  

The solution of the Schr\"odinger equation for the vibrational motion with the  potential of Eq. (\ref{eq:couple_pots}) is a function $\Psi$ having two components $\psi_\pm$ corresponding to the two diabatic electronic states. In the adiabatic basis, corresponding to the $\vert l\rangle$ and $\vert u\rangle$   electronic states, the two components $\psi_{l,u}$ of $\Psi$ have the form
\begin{eqnarray}
\label{eq:geometrical_phase}
\psi_{l} = (\psi_- e^{if/2} + \psi_+e ^{-if/2})/\sqrt{2}\,,\nonumber\\
\psi_{u} = i(\psi_- e^{if/2} - \psi_+e ^{-if/2})/\sqrt{2}\,.
\end{eqnarray}

The $C_3$ symmetry operator applied three times  to the molecule returns the three nuclei back to their original relative configuration. However the components $\psi_{l,u}$ change sign  $\psi_{l,u}\to-\psi_{l,u}$ under the $C_3^3$ operator because $f\to f+2\pi$. It is a well-known property of adiabatic states when there is a conical intersection. The property is often referred to as geometrical or Berry phase effect. Because the adiabatic electronic wave functions $\vert l\rangle$ and $\vert u\rangle$ also change sign, the total vibronic wave function 
\begin{eqnarray}
\label{eq:ad_di_wf}
\Psi = \psi_{l} \vert l\rangle +\psi_{u} \vert u\rangle =\psi_{+} \vert +\rangle +\psi_- \vert -\rangle
\end{eqnarray}
is unchanged after the $C_3^3$ operation. 

Therefore, the $\psi_{l,u}$  components are two-valued. This complicates the solution of the vibrational Schr\"odinger equation in the adiabatic basis. In this situation, the diabatic basis is preferable because the diabatic components are single-valued. In addition, in the diabatic representation, there is no need to deal with first and second derivatives of electronic wave functions in order to account for non-adiabatic couplings between the two adiabatic PES.

\subsection{Nuclear dynamics}
The hyperspherical coordinates  $R,\theta,\phi$ are specially adapted to solve the vibrational Schr\"odinger equation for \Http. Solving the Schr\"odinger equation, we use the two-step approach \cite{kokoouline05}. First, we solve the equation for a fixed hyper-radius $R_j$
\begin{equation}
\label{eq:Had}
H_{R_j} \varphi_{j,a}(\theta,\phi)=U_a(R_j)\varphi_{j,a}(\theta,\phi)\,,
\end{equation}
where $H_{R_j} $ is the two-channel vibrational Hamiltonian with the hyper-radius fixed at $R_j$. The solution of the above eigenvalue problem gives eigenstates  $\varphi_{j,a}(\theta,\phi)$ depending on the two hyperangles, which are usually called hyperspherical adiabatic (HSA) states, and eigenvalues $U_a(R_j)$, called HSA potentials.   

Once the HSA states are determined, the complete vibrational wave function $\psi(R,\theta,\phi)$ of the two-component triplet potential 
is represented as an expansion $\psi(R,\theta,\phi)=\sum_{j,a}y_{j,a}(R,\theta,\phi)c_{j a}$ in the basis of following functions 
$y_{j,a}=\pi_j(R)\varphi_{j,a}(\theta,\phi)$, where $\varphi_{j,a}(\theta,\phi)$ are calculated at the first step, $\pi_j(R)$ are DVR basis functions along the hyper-radius, localized at values $R_j$ of the hyper-radius. Coefficients $c_{j a}$ are unknown. The expansion functions $y_{j,a}$ are not necessarily  orthogonal to each other because HSA states $\varphi_{j,a}(\theta,\phi)$ evaluated at different $R_j$ are not orthogonal. Therefore, in the basis $y_{j,a}$, the vibrational Schr\"odinger equation for $\psi(R,\theta,\phi)$ takes the form of a generalized eigenvalue problem for coefficients 
$c_{j a}$
\begin{eqnarray}
 \label{eq:gen_eigen}
 \sum_{j',a'}\Big[\langle\pi_{j} |-\frac{\hbar^2}{2\mu}\frac{d^2}{dR^2}|\pi_{j'}\rangle{\cal O}_{ja,j'a'}+\langle\pi_{j}|U_a(R)|\pi_{j'}\rangle\delta_{aa'}\Big]c_{j'a'} \nonumber\\ 
=E_n ^{vib} \sum_{j',a'}\langle\pi_{j}|\pi_{j'}\rangle{\cal O}_{ja,j'a'}c_{j'a'}\,,
\end{eqnarray}
where $\mu$ is the three-body reduced mass of the ion; ${\cal O}_{ja,j'a'}$ are overlaps between HSA states 
\begin{equation}
\label{eq:couplings_svd}
 {\cal O}_{ja,j'a'}=\langle\varphi_{j,a}(\theta,\phi)|\varphi_{j'a'}(\theta,\phi)\rangle\,.
\end{equation}

\section{Bound states of the H$_3^+$ triplet state}
\label{sec:bound}

At equilibrium, the \Http molecule in the triplet state is linear. Three equivalent minima exist due
to permutationial symmetry of identical nuclei, separated by barriers. Since the barriers are not too high,
tunneling is a feasible operation. Furthermore, most of the vibrational states are located above the barriers. 
Symmetry classification of the rotation-vibrational states of this floppy molecule  then requires
the use of the full three-particle permutation-inversion group $S_3 \times I$, which is isomorphic with the $D_{3h}(M)$ 
molecular symmetry group. 
Pure vibrational states ($J=0$) must have even parity and hence transform as one of the three irreducible representations
of the subgroup $C_{3v}(M)$, $A_1'$, $A_2'$, and $E'$. Below, we will omit the primes. 

To obtain vibrational energies and wave functions, first, we have obtained the HSA wave functions and energies. In the numerical calculations, we used B-splines \cite{deboor01} to solve the hyperangular part of the Schr\"odinger equation. In a typical calculation, we used 180 B-spline intervals along angle $\theta$, which changes in the interval $[-\pi,\pi]$. The angle $\phi$ changes in the interval $[0,2\pi]$. However, because of the symmetry of the potential as a function of $\phi$, each HSA state $\varphi$ (as well as the total vibrational functions $\Psi$) belongs to one of the three irreps $A_1$, $A_2$, and $E$ and, as a result, has a certain periodicity as a function of $\phi$. The periodicity allows us to reduce the $[0,2\pi]$ interval: To obtain HSA states of the $A_1$ and $A_2$ irreps, the interval  $[-\pi/2,\pi/6]$ is sufficient, and for states  corresponding to the $E$ irrep, the interval $[-\pi/2,\pi/2]$ is enough, provided that appropriate boundary conditions are applied to the HSA states at ends of the intervals. For the $A_1$ states and one ($E_a$) of the two components of the $E$ irrep, derivative of wave functions with respect to $\phi$ are zero at the ends. For the $A_2$ states and the second component ($E_b$) of the $E$ irrep, the wave functions itself are zero at the ends. In a typical calculation, we use 100 B-splines along the $\phi$ to obtain $A_1$ and $A_2$ states and 211 B-splines  to obtain $E$ states.

\begin{figure}
\includegraphics[height=16cm]{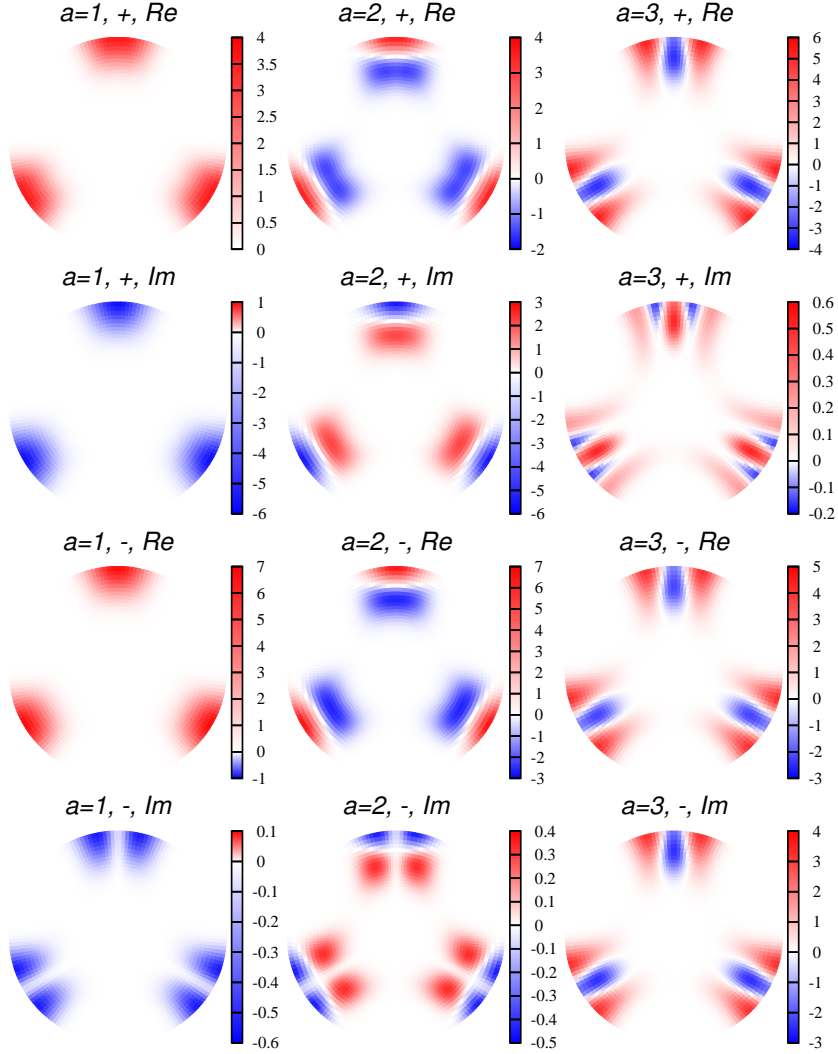}
\caption{The first three ($a=1,2,3$) HSA wave functions $\varphi_{j,a}(\theta,\phi)$ of the $A_1$ irreducible representation of \Http in the \textit{diabatic} representation obtained for a fixed value of hyper-radius $R_j=4.7 a_0$. Each HSA wave function has two complex-valued components, $\varphi_+$ and $\varphi_-$ similarly to two functions $\psi_+$ and $\psi_-$ of Eq.~(\ref{eq:ad_di_wf}). They are shown as a column of four graphs for each $a$.}
\label{fig:wfd_H3_A_1}
\end{figure}

\begin{figure}
\includegraphics[height=16cm]{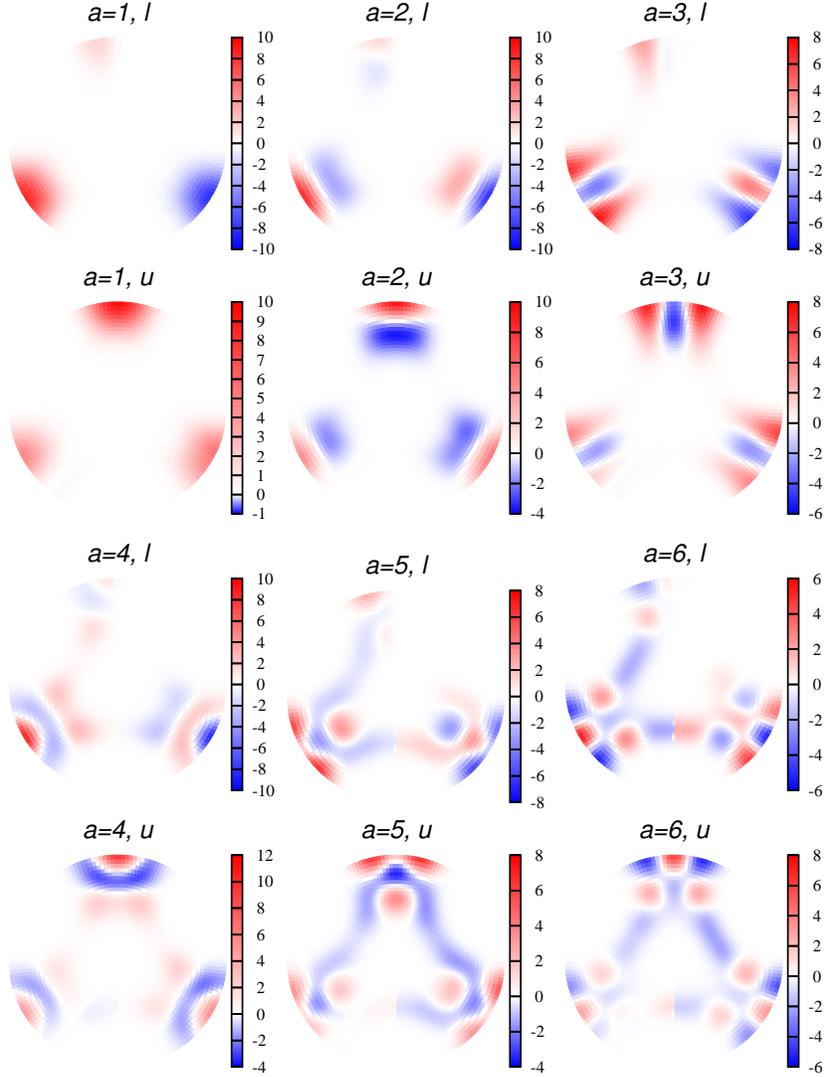}
\caption{The  first six ($a=1,\cdots,6$) $A_1$ HSA wave functions $\varphi_{j,a}(\theta,\phi)$ shown here in the \textit{adiabatic} representation. The first three functions are the same as in Fig. \ref{fig:wfd_H3_A_1}. In the adiabatic representation, the wave functions could be made real and, therefore, the imaginary parts are zero and are not shown here. Notice that the two adiabatic components of the vibrational  wave functions change sign when the hyperspherical  angle $\phi$ goes through value $-\pi/2$, which is in agreement with to Eq.~(\ref{eq:geometrical_phase}). Also, notice that the white color corresponds to $\varphi_{j,a}=0$ in all graphs, but the correspondence of intensity of red and blue colors to actual values of $\varphi_{j,a}$ is slightly different in the graphs.}
\label{fig:wfa_H3_A_1}
\end{figure}

\begin{figure}
\includegraphics[height=16cm]{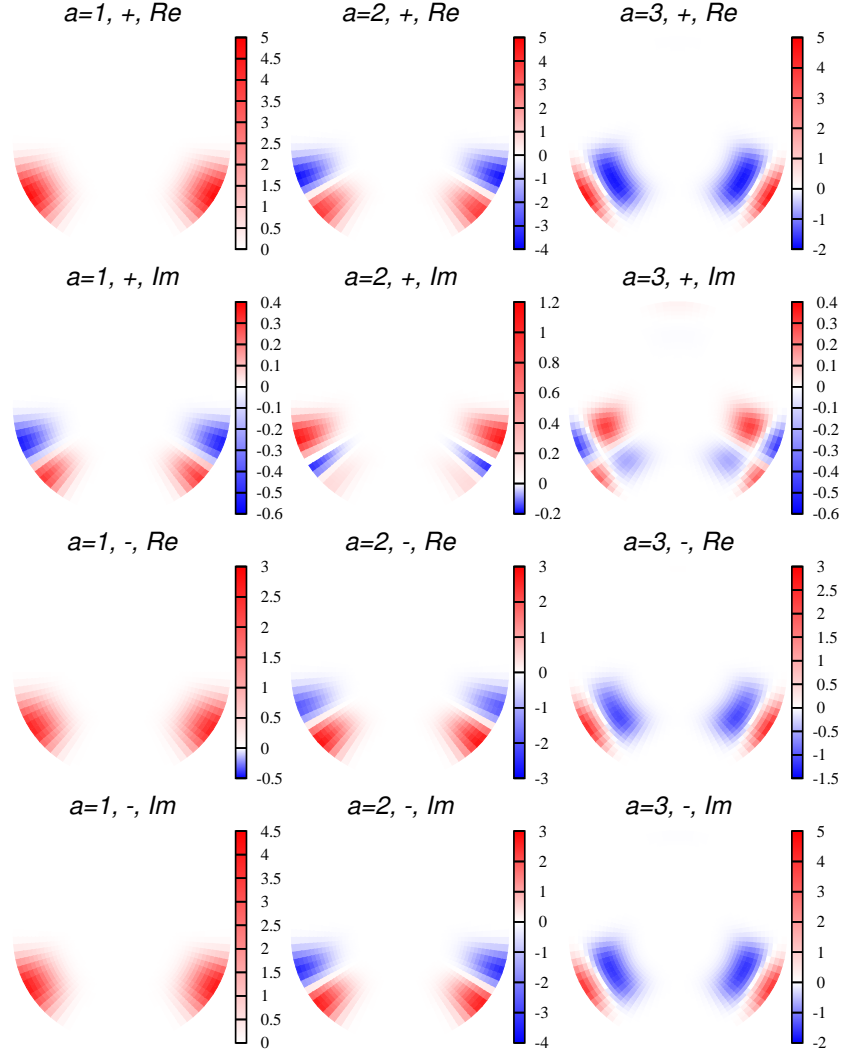}
\caption{The first three ($a=1,2,3$) HSA wave functions $\varphi_{j,a}(\theta,\phi)$ of the $E_a$ irreducible representation in the \textit{diabatic} representation calculated for $R_j=4.7 a_0$. Each HSA wave function has two complex-valued components, $\varphi_+$ and $\varphi_-$ similarly to two functions $\psi_+$ and $\psi_-$ of Eq.~(\ref{eq:ad_di_wf}). They are shown as a column of four graphs for each $a$.}
\label{fig:wfd_H3_E_a}
\end{figure}

\begin{figure}
\includegraphics[height=16cm]{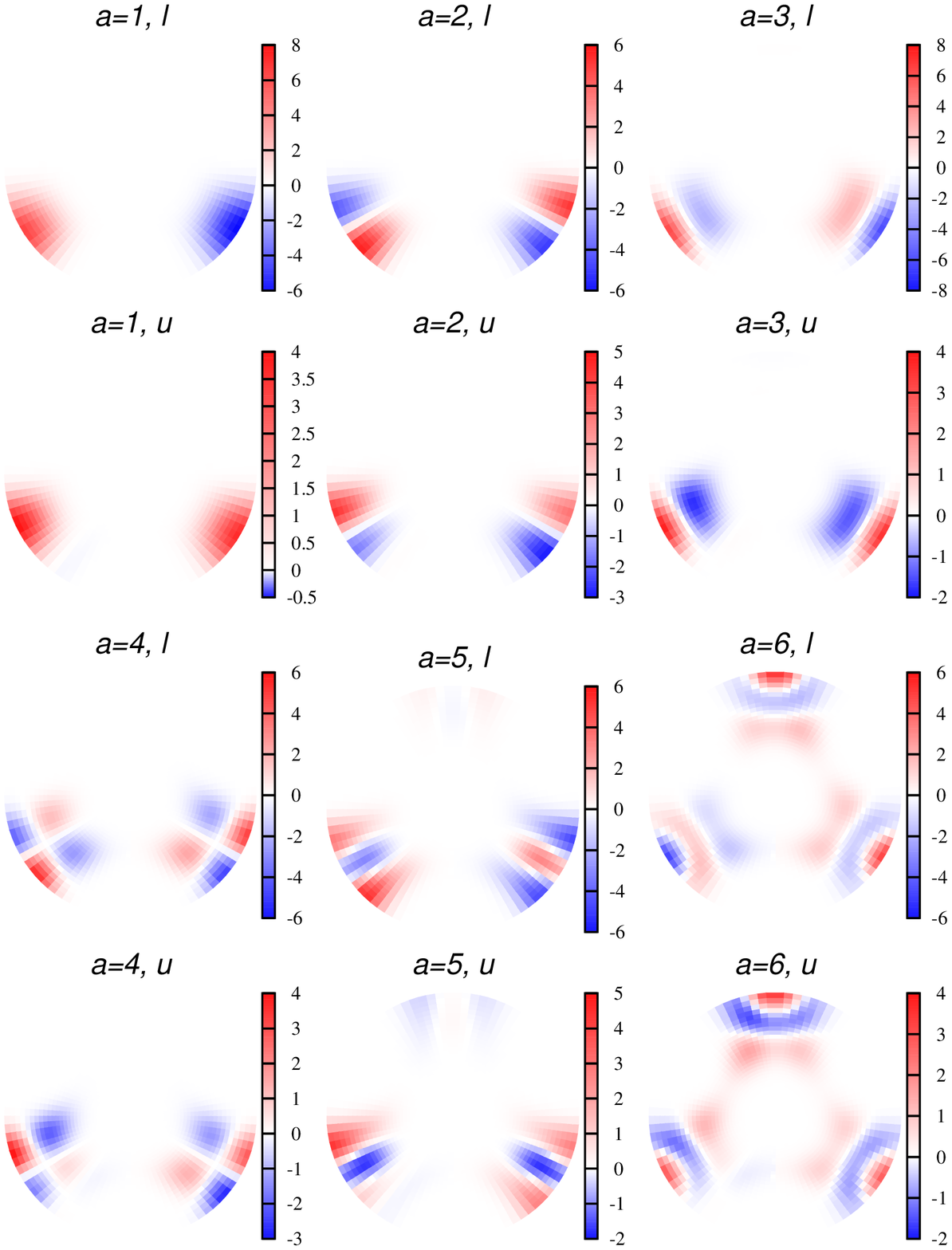}
\caption{The  first six ($a=1,\cdots,6$)  $E_a$ HSA wave functions $\varphi_{j,a}(\theta,\phi)$ shown here in the \textit{adiabatic} representation. The first three functions are the same as in Fig. \ref{fig:wfd_H3_E_a}. }
\label{fig:wfa_H3_E_a}
\end{figure}

The wave functions $\Psi(R,\theta,\phi)$ and $\varphi_{j,a}(\theta,\phi)$ have two electronic components. By definition, the electronic diabatic  states $\vert +\rangle$ and $\vert -\rangle$ transform as components $E_+$ and $E_-$ of the $E$ irrep. Therefore, the irrep $\Gamma_{vibr}$  
of the vibronic function is related to the irrep  $\Gamma_{vib}$ of the vibrational components $\psi_{\pm}$ as $\Gamma_{vibr}=\Gamma_{vib}\times E$.
The vibrational adiabatic components  $\varphi_{j,a,e}(\theta,\phi)$ with $e=l$ or $u$ exhibit the property of geometrical phase: If the geometry of three nuclei is changing continuously in such a way that it makes a closed loop in the space of geometries and the equilateral configuration stays inside the loop, the wave function changes sign when the geometry returns to the initial one. Therefore, adiabatic components of total wave functions could be considered as two-valued: An arbitrary configuration of the \Http nuclei described by an angle $\alpha$ of distortion (or by a hyperangle $\phi$) can equally be described by $\alpha+2\pi$ (or $\phi+2\pi$). Each adiabatic component is then two-valued $\varphi_{j,a,e}(\theta,\phi)=-\varphi_{j,a,e}(\theta,\phi+2\pi)$. 
Because the electronic adiabatic states are also two-valued, the total wave 
function stays single-valued.

Examples of diabatic and adiabatic vibrational components of  $\varphi_{j,a}(\theta,\phi)$ are shown in Figs. \ref{fig:wfd_H3_A_1} and  \ref{fig:wfa_H3_A_1} for a fixed value of $R=4.7\ a_0$ for $\Gamma_{vib}=A_1$. Figures \ref{fig:wfd_H3_E_a} and  \ref{fig:wfa_H3_E_a} shows several HSA states for $\Gamma_{vib}=E$. Wave functions of bound states in the diabatic representation of complex-values electronic states $\vert +\rangle$ and $\vert -\rangle$ are in general also complex. However, adiabatic states  $\vert l\rangle$ and $\vert u\rangle$ are read and, therefore, the overall phase of the adiabatic vibrational components could be made real. In Figs. \ref{fig:wfd_H3_A_1} and \ref{fig:wfd_H3_E_a}  we show real and imaginary parts of the two components of lowest three ($a=1,2,3$)  HSA  $\varphi_{j,a}(\theta,\phi)$ in the diabatic representation. Figures   \ref{fig:wfa_H3_A_1} and  \ref{fig:wfa_H3_E_a} show only real part of the two components of lowest six ($a=1,\cdots,6$) HSA. The imaginary part of the adiabatic components is zero. 

The property geometrical phase is demonstrated by vibrational adiabatic components of $\Gamma_{vib}=A_1$ and $E$ in Figs.  \ref{fig:wfa_H3_A_1} and \ref{fig:wfa_H3_E_a}. Only one sheet of the two-valued functions is shown. The equilateral geometry is the branch point for the cut. The configuration space of two hyperangle is cut from the branch point straight down to the linear geometry along the line  $\phi=-\pi/2$ or $\alpha=\pi$. The choice of the cut is, of cause, arbitrary as long as it goes from the branch point to a linear geometry. As it is demonstrated by Figs.  \ref{fig:wfa_H3_A_1} and \ref{fig:wfa_H3_E_a}, the adiabatic components indeed change sign at  the cut. In Fig. \ref{fig:wfa_H3_E_a} it is less evident, except $a=6$, because the wave functions with $a=1,\cdots,5$ are almost zero near the cut.

\begin{figure}
\includegraphics[width=10cm]{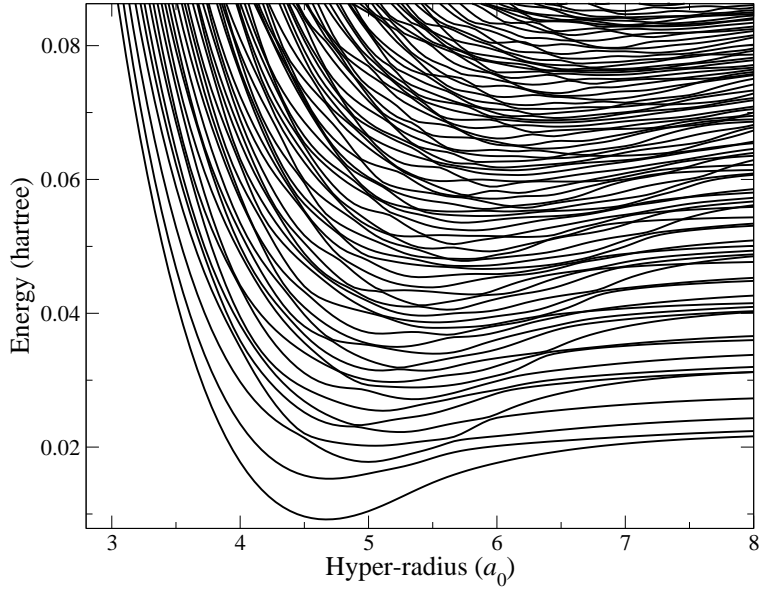}
\caption{Hyperspherical adiabatic  potentials of $A_1$ vibrational symmetry obtained for the triplet state of \Http. Due to the fact that the conical intersection between the surfaces of the two components of the triplet state is situated at a relatively high energy compared to the minimum of the potential, the difference between the HSA potentials obtained in one- and two-state calculations is small, about $3\times10^{-5} \, \Eh$,  and cannot be seen at the figure. }
\label{fig:adiabatic_energies_A1}
\end{figure}

To assess the effect of the coupling with the upper PES, we have made two types of calculations. First, we determined HSA states and vibrational energies using only the lowest adiabatic potential $V_l$ -- the one-state calculation. In the second type of calculation, we solve the Schr\"odinger equation for the coupled potential of Eq.~(\ref{eq:couple_pots}) -- the two-state calculation. The difference between energies of obtained vibrational states reflects the effect of the coupling between the two adiabatic surfaces on the vibrational energies of $V_l$.  Figure \ref{fig:adiabatic_energies_A1} shows the HSA energies $U_a(R)$ obtained in one- and two-state calculations. The HSA potentials curves are very similar to each other in one- and two-state calculations in the energy region, shown in the figure. For example, near the minimum of the lowest HSA curve, the energy difference between the two calculations is about $3\times 10^{-5} \, \Eh$.

Once the HSA energies and wave functions are calculated, we solve the hyper-radial part (Eq.~\ref{eq:couplings_svd})  
of the Schr\"odinger equation. Energies of obtained vibrational levels in one- and two-state calculations are listed in 
Table~\ref{tab:H3p_energies}. In the table, symbols of irreducible representations refer to symmetry of the vibrational part 
in one-state calculation, and to the symmetry of the diabatic vibrational components in the two-state calculation. 
In addition, we provide an identification of the vibrational states in terms of $D_{\infty h}(M)$ linear molecule spectroscopic quantum numbers  
$(v_1, v_2^{|l|}, v_3)$, where $v_1$ and $v_3$ correspond to the symmetric and antisymmetric stretching vibrations, 
and $v_2$ to the degenerate bending vibration, with $l$ the associated vibrational angular momentum.
Such a classification is not exact, since the molecule is not localized in a single potential well.
Rather, linear combinations of the three equivalent vibrational wavefunctions must be formed, which then 
give rise to a one-dimensional and a two-dimensional representation in the $C_{3v}(M)$ group.
For not too highly excited vibrational states, the linear molecule quantum numbers could be
useful to label the vibrational states if supplemented by a degeneracy index ($A_1/A_2$ or $E$).
A discussion of the quantum numbers can be found in Refs.~\cite{FRI01:1183,ALI03:163}.
Results of the one-state calculation may be compared with that a previous calculation~\cite{varandas05}, 
where the same surface $V_l$ was used. The results agree well except for the highest states, 
which in Ref.~\cite{varandas05} could not be converged to the same accuracy as the lower states.

Comparing now the relative energies of the one-state and two-state calculations, they agree to within
$\approx 1 \, \cm$ for levels up to $\approx 2300 \, \cm$. 
These small differences can be attributed to the neglect of non-adiabatic effects in the one-state calculation.
It is worth noting that only the first three states are located below the electronic energy of the
$\rm \Htp + H$ channel ($E = 1228.5 \, \cm$), while the large remainder lies between this threshold and
the vibrational zero point energy of \Htp. 

The splitting in energy between the one-dimensional and the two-dimensional components of a linear molecule type vibrational 
state is expected to be small for the lowest energy levels because the tunneling is weak. For more excited vibrational levels the 
splitting is expected to increase, which is indeed observed in one- and two-state calculations as evident from Table~\ref{tab:H3p_energies}. 
The tunneling between the three potential wells is more effective in the two-state calculations. It happens because the non-adiabatic 
coupling with the upper potential surface $V_u$ makes tunneling easier if compared to the calculation involving only the $V_l$ surface. 
For example, the lowest excited state $(0,0^0,0)E$ in one-state calculation is just about $2\times 10^{-4}$cm$^{-1}$ 
above the ground vibrational level $(0,0^0,0)A_1$. In the two-state calculation, the energy difference between 
$(0,0^0,0)E$ and  $(0,0^0,0)A_1$ is 1.04~\cm. For the second state, $(0,0^0,1)$, the splitting is 0.05~\cm in the
two-state calculation. The higher vibrational states are already located above the barriers.

\begin{sidewaystable}
\caption{Energies of vibrational levels obtained in one- and two-state calculations. The origin is chosen at the energy of the ground vibrational level $ (0,0^0,0) $. In one-state calculation, the zero-point energy is  $E_0=1722.043 \, \cm$, the energy of H$_2^+(v=0,j=0)$+H dissociation is $E_{thres}= 2382.0\ \cm$ above $(0,0^0,0)$. 
}
\label{tab:H3p_energies}
\vspace{-1cm}
\begin{tabular}{c rr rr rr rr rr rr}
\\
\\
 & \multicolumn{6}{c}{1-state calculation} & \multicolumn{6}{c}{2-state calculation}  \\
$(v_1,v_2^{\ell},v_3)$ & $i$ & $A_1^{\prime}$ & $i$ & $A_2^{\prime}$ & $i$ & $E^{\prime}$ & $i$ & $A_1^{\prime}$ & $i$ & $A_2^{\prime}$ & $i$ & $E^{\prime}$ \\
\hline
\hline
  $ (0,0^0,0) $ & ~~0 &         0.00 &     &              & ~~0 &         0.00 & ~~0 &         0.00 &     &              & ~~0 &         1.04 \\
  $ (0,0^0,1) $ &     &              & ~~0 &       738.49 &   1 &       738.49 &     &              & ~~0 &       738.91 &   1 &       738.86 \\
  $ (1,0^0,0) $ &   1 &       975.05 &     &              &   2 &       975.06 &   1 &       975.34 &     &              &   2 &       976.07 \\
  $ (0,2^0,0) $ &   2 &      1273.73 &     &              &   3 &      1273.79 &   2 &      1274.13 &     &              &   3 &      1275.01 \\
  $ (1,0^0,1) $ &     &              &   1 &      1474.57 &   4 &      1474.4  &     &              &   1 &      1475.24 &   4 &      1475.20 \\
  $ (0,0^0,2) $ &   3 &      1573.70 &     &              &   5 &      1573.8  &   3 &      1574.17 &     &              &   5 &      1574.96 \\
  $ (0,2^0,1) $ &     &              &   2 &      1730.39 &   6 &      1728.4  &     &              &   2 &      1730.77 &   6 &      1728.9  \\
  $ (2,0^0,0) $ &   4 &      1922.55 &     &              &   7 &      1923.1  &   4 &      1922.94 &     &              &   7 &      1924.0 \\
  $ (1,2^0,0) $ &   5 &      1940.34 &   3 &              &   8 &      1951.0  &   5 &      1940.88 &   3 &              &   8 &      1951.9 \\
  $ (0,0^0,3) $ &     &              &   4 &      1972.87 &   9 &      1970.7  &     &              &   4 &      1974.16 &   9 &      1972.2 \\
  $ (1,0^0,2) $ &   6 &      2158.72 &     &              &  10 &      2136.9  &   6 &      2159.79 &     &              &  10 &      2137.9 \\
  $ (0,4^0,0) $ &   7 &      2188.49 &     &              &  11 &      2166.4  &   7 &      2189.69 &     &              &  11 &      2167.9 \\
  $ (2,0^0,1) $ &     &              &   5 &      2205.01 &  12 &      2251.0  &     &              &   5 &      2205.79 &  12 &      2253.5 \\
  $ (1,2^0,1) $ &     &              &   6 &      2271.13 &  13 &      2259.3  &     &              &   6 &      2273.61 &  13 &      2261.2 \\
  $ (0,2^0,2) $ &   8 &      2308.61 &     &              &  14 &      2334.9  &   8 &      2311.47 &     &              &  14 &      2338.1 \\
  $ (3,0^0,0) $ &   9 &      2340.09 &     &              &  15 &      2357.   &   9 &      2341.43 &     &              &  15 &      2358.6 \\
  $           $ &   10&      2365.87 &     &              &     &              &  10 &      2371.42 &     &              &  16 &       2362.9\\
  $           $ &   11&      2372.72 &     &              &     &              &  11 &      2381.   &     &              &  17 &      2379.9 \\
  $           $ &     &              &   7 &      2402.28 &     &              &     &              &   7 &      2407.00 &     &             \\
\hline
\hline
\end{tabular}
\end{sidewaystable}

Not all of the computed states shown in Table~\ref{tab:H3p_energies} are allowed for this system with three
identical protons. Since protons are fermions, the total wavefunction must be antisymmetric under permutation of the
position of two nuclei and hence transform as $A_2^{\prime}$ or $A_2^{\prime \prime}$ in the group $D_{3h}(M)$.
The total wavefunction $\Psi_{tot}$ has contributions from the nuclear spin ($ns$), electronic spin ($es$) and the
spatial parts, electronic ($el$), vibrational ($vib$) and rotational ($rot$). If rotation is neglected ($J=0$),
classification in the $C_{3v}(M)$ symmetry group is sufficient. The electronic spin function is totally symmetric and transforms as
$A_1$. It suffices to analyse the nuclear spin, electronic and vibrational symmetries.

Two nuclear spin functions may be formed, one having $I=3/2$ (ortho spin) and the other $I=1/2$ (para spin). 
They transform as $A_1$ and $E$, respectively. The symmetry of the electronic functions in the adiabatic
representation is $A_2$ for the lower surface and $A_1$ for the upper, while in the diabatic representation
they transform as two components, $E_+$ and $E_-$ (or $E_a$ and $E_b$), of the degenerate representation $E$. Hence, with respect
to their symmetry under the $(12)$ permutation operation, the lower adiabatic state is antisymmetric and the 
upper symmetric. This is different for the diabatic representation, where the two components are
transformed between each other, see Eq.~(\ref{eq:Eplus_minus}).

In the adiabatic representation, we have for ortho nuclear spin the direct  product
$\Gamma_{el} \times \Gamma_{ns} = A_2 \times A_1 = A_2$
which requires the vibrational state to have $A_1$ symmetry so that the total function
transfroms as
\begin{equation}
\Gamma_{tot} = \Gamma_{el} \times \Gamma_{ns} \times \Gamma_{vib} = A_2 \times A_1 \times A_1 = A_2
\end{equation}
For para spin, $\Gamma_{el} \times \Gamma_{ns} = A_2 \times E = E$, which is compatible with $E$ vibrational symmetry, since
\begin{equation}
\Gamma_{tot} = \Gamma_{el} \times \Gamma_{ns} \times \Gamma_{vib} = A_2 \times E \times E = A_2 \left( + A_1 + E \right)\,,
\end{equation}
where the $A_1$ and $E$ irreps are taken in parentheses because they are not allowed for the complete wave functions of H$_3^+$.

We now turn to the diabatic representation, where we have $\Gamma_{el} = E$. Hence for ortho nuclear spin
$\Gamma_{el} \times \Gamma_{ns} = E \times A_1 = E$ which selects $\Gamma_{vib}=E$, since
\begin{equation}
\Gamma_{tot} = \Gamma_{el} \times \Gamma_{ns} \times \Gamma_{vib} = E \times A_1 \times E = A_2 \left( + A_1 + E \right)
\end{equation}

For para nuclear spin we obtain
$\Gamma_{el} \times \Gamma_{ns} = E \times E = A_1 + A_2 + E$
which may be combined with any vibrational symmetry, 
\begin{equation}
\Gamma_{tot} = \Gamma_{el} \times \Gamma_{ns} \times \Gamma_{vib} = 
\left\{ \begin{array}{lcl}
E \times E \times A_1 & = & A_2  \left( + A_1 + E \right) \\
E \times E \times A_2 & = & A_2  \left( + A_1 + E \right) \\
E \times E \times E   & = & A_2  \left( + A_1 + 3 E \right) \\
\end{array} \right.
\end{equation}

An important consequence of the geometrical phase is therefore its alteration
of the spin statistical weights of the vibrational functions which we present in Table~\ref{tab:H3p_stat_weights}.

\begin{table}
\caption{Nuclear spin statistical weights of the vibrational wave functions obtained
in a one-state and a two-state calculation. The geometrical phase present in the two-state calculation
alters the spin statistical weights. In particular, it makes $A_2$ states allowed. This difference could be
crucial to diagnose the geometrical phase experimentally.}
\label{tab:H3p_stat_weights}
\begin{tabular}{ccc}
\hline
$\Gamma_{vib}$ &  $W(1\, st.)$   & $W(2\, st.)$     \\
\hline
\hline
$A_1$  &  4     &  2            \\
$A_2$  &  0     &  2            \\
$E$    &  2     &  4            \\
\hline
\hline
\end{tabular}
\end{table}

\section{Conclusion}

Vibrational energies and wave functions  of the triplet state of the \Http ion have been determined. In the calculations, 
the ground and first excited triplet electronic states are included as well as the non-adiabatic coupling between them. 
The calculations have been performed in a diabatic basis using a diabatization procedure suggested by Longuet-Higgins transforming  
two adiabatic {\it ab initio} potential energy surfaces of the  triplet-\Http state into a $2\times2$ matrix. The suggested 
theoretical approach allows us to account for the non-adiabatic coupling between the ground and the first excited electronic states of 
triplet-\Http. It also includes the effect of the geometrical phase due to the conical intersection between the two adiabatic potential surfaces. 
The results are compared to the calculation involving only the lowest adiabatic potential energy surface of the triplet-\Http ion 
and neglecting the geometrical phase. The energy difference between results with and without the non-adiabatic coupling 
and the geometrical phase is about one wave number for the lowest vibrational levels. However, in the presence of the
geometrical phase the spin statistical weights of the vibrational states are altered. These findings are consistent with the
expected  small effect on the vibrational energy levels as the excited triplet state is located far
in the continuum of the lower state. They are also consistent with the fact that the geometrical phase is topological
and therefore manifests itself even if the two electronic states are well separated in energy.

\section*{Acknowledgments}
This work is supported by the Conseil Regional de la Region Champagne-Ardenne, 
(ESRI/Sdel/OD-20130604) the National Science Foundation (Grant No PHY-10-68785) 
and the ROMEO HPC Center at the University of Reims Champagne-Ardenne.


\begin{thebibliography}{17}
\expandafter\ifx\csname natexlab\endcsname\relax\def\natexlab#1{#1}\fi
\expandafter\ifx\csname bibnamefont\endcsname\relax
  \def\bibnamefont#1{#1}\fi
\expandafter\ifx\csname bibfnamefont\endcsname\relax
  \def\bibfnamefont#1{#1}\fi
\expandafter\ifx\csname citenamefont\endcsname\relax
  \def\citenamefont#1{#1}\fi
\expandafter\ifx\csname url\endcsname\relax
  \def\url#1{\texttt{#1}}\fi
\expandafter\ifx\csname urlprefix\endcsname\relax\def\urlprefix{URL }\fi
\providecommand{\bibinfo}[2]{#2}
\providecommand{\eprint}[2][]{\url{#2}}

\bibitem[{\citenamefont{Oka}(2006)}]{oka06b}
\bibinfo{author}{\bibfnamefont{T.}~\bibnamefont{Oka}}, \bibinfo{journal}{Proc.
  Nat. Ac. Scien.} \textbf{\bibinfo{volume}{103}}, \bibinfo{pages}{12235}
  (\bibinfo{year}{2006}).

\bibitem[{\citenamefont{Oka}(2013)}]{OKA13:8738}
\bibinfo{author}{\bibfnamefont{T.}~\bibnamefont{Oka}}, \bibinfo{journal}{Chem.
  Rev.} \textbf{\bibinfo{volume}{113}}, \bibinfo{pages}{8738}
  (\bibinfo{year}{2013}).

\bibitem[{\citenamefont{Schaad and Hicks}(1974)}]{SCH74:1934}
\bibinfo{author}{\bibfnamefont{L.~J.} \bibnamefont{Schaad}} \bibnamefont{and}
  \bibinfo{author}{\bibfnamefont{W.~V.} \bibnamefont{Hicks}},
  \bibinfo{journal}{J. Chem. Phys.} \textbf{\bibinfo{volume}{61}},
  \bibinfo{pages}{1934} (\bibinfo{year}{1974}).

\bibitem[{\citenamefont{Friedrich et~al.}(2001)\citenamefont{Friedrich, Alijah,
  Xu, and Varandas}}]{FRI01:1183}
\bibinfo{author}{\bibfnamefont{O.}~\bibnamefont{Friedrich}},
  \bibinfo{author}{\bibfnamefont{A.}~\bibnamefont{Alijah}},
  \bibinfo{author}{\bibfnamefont{Z.~R.} \bibnamefont{Xu}}, \bibnamefont{and}
  \bibinfo{author}{\bibfnamefont{A.~J.~C.} \bibnamefont{Varandas}},
  \bibinfo{journal}{Phys. Rev. Lett.} \textbf{\bibinfo{volume}{86}},
  \bibinfo{pages}{1183} (\bibinfo{year}{2001}).

\bibitem[{\citenamefont{Sanz et~al.}(2001)\citenamefont{Sanz, Roncero, Tablero,
  Aguado, and Paniagua}}]{SAN01:2182}
\bibinfo{author}{\bibfnamefont{C.}~\bibnamefont{Sanz}},
  \bibinfo{author}{\bibfnamefont{O.}~\bibnamefont{Roncero}},
  \bibinfo{author}{\bibfnamefont{C.}~\bibnamefont{Tablero}},
  \bibinfo{author}{\bibfnamefont{A.}~\bibnamefont{Aguado}}, \bibnamefont{and}
  \bibinfo{author}{\bibfnamefont{M.}~\bibnamefont{Paniagua}},
  \bibinfo{journal}{J. Chem. Phys.} \textbf{\bibinfo{volume}{114}},
  \bibinfo{pages}{2182} (\bibinfo{year}{2001}).

\bibitem[{\citenamefont{Jahn and Teller}(1937)}]{JAH37:220}
\bibinfo{author}{\bibfnamefont{H.~A.} \bibnamefont{Jahn}} \bibnamefont{and}
  \bibinfo{author}{\bibfnamefont{E.}~\bibnamefont{Teller}},
  \bibinfo{journal}{Proc. R. Soc.} \textbf{\bibinfo{volume}{161A}},
  \bibinfo{pages}{220} (\bibinfo{year}{1937}).

\bibitem[{\citenamefont{Cernei et~al.}(2003)\citenamefont{Cernei, Alijah, and
  Varandas}}]{CER03:2637}
\bibinfo{author}{\bibfnamefont{M.}~\bibnamefont{Cernei}},
  \bibinfo{author}{\bibfnamefont{A.}~\bibnamefont{Alijah}}, \bibnamefont{and}
  \bibinfo{author}{\bibfnamefont{A.~J.~C.} \bibnamefont{Varandas}},
  \bibinfo{journal}{J. Chem. Phys.} \textbf{\bibinfo{volume}{118}},
  \bibinfo{pages}{2637} (\bibinfo{year}{2003}).

\bibitem[{\citenamefont{Alijah et~al.}(2003)\citenamefont{Alijah, Viegas,
  Cernei, and Varandas}}]{ALI03:163}
\bibinfo{author}{\bibfnamefont{A.}~\bibnamefont{Alijah}},
  \bibinfo{author}{\bibfnamefont{L.~P.} \bibnamefont{Viegas}},
  \bibinfo{author}{\bibfnamefont{M.}~\bibnamefont{Cernei}}, \bibnamefont{and}
  \bibinfo{author}{\bibfnamefont{A.~J.~C.} \bibnamefont{Varandas}},
  \bibinfo{journal}{J. Mol. Spectrosc.} \textbf{\bibinfo{volume}{221}},
  \bibinfo{pages}{163} (\bibinfo{year}{2003}).

\bibitem[{\citenamefont{Viegas et~al.}(2004)\citenamefont{Viegas, Cernei,
  Alijah, and Varandas}}]{VIE04:253}
\bibinfo{author}{\bibfnamefont{L.~P.} \bibnamefont{Viegas}},
  \bibinfo{author}{\bibfnamefont{M.}~\bibnamefont{Cernei}},
  \bibinfo{author}{\bibfnamefont{A.}~\bibnamefont{Alijah}}, \bibnamefont{and}
  \bibinfo{author}{\bibfnamefont{A.~J.~C.} \bibnamefont{Varandas}},
  \bibinfo{journal}{J. Chem. Phys.} \textbf{\bibinfo{volume}{120}},
  \bibinfo{pages}{253} (\bibinfo{year}{2004}).

\bibitem[{\citenamefont{Varandas et~al.}(2005)\citenamefont{Varandas, Alijah,
  and Cernei}}]{varandas05}
\bibinfo{author}{\bibfnamefont{A.~J.} \bibnamefont{Varandas}},
  \bibinfo{author}{\bibfnamefont{A.}~\bibnamefont{Alijah}}, \bibnamefont{and}
  \bibinfo{author}{\bibfnamefont{M.}~\bibnamefont{Cernei}},
  \bibinfo{journal}{Chem. Phys.} \textbf{\bibinfo{volume}{308}},
  \bibinfo{pages}{285} (\bibinfo{year}{2005}).

\bibitem[{\citenamefont{Alijah and Varandas}(2006)}]{ALI06:2889}
\bibinfo{author}{\bibfnamefont{A.}~\bibnamefont{Alijah}} \bibnamefont{and}
  \bibinfo{author}{\bibfnamefont{A.~J.~C.} \bibnamefont{Varandas}},
  \bibinfo{journal}{Philos. Trans. R. Soc. London A}
  \textbf{\bibinfo{volume}{364}}, \bibinfo{pages}{2889} (\bibinfo{year}{2006}).

\bibitem[{\citenamefont{Blandon and Kokoouline}(2009)}]{blandon09}
\bibinfo{author}{\bibfnamefont{J.}~\bibnamefont{Blandon}} \bibnamefont{and}
  \bibinfo{author}{\bibfnamefont{V.}~\bibnamefont{Kokoouline}},
  \bibinfo{journal}{Phys. Rev. Lett.} \textbf{\bibinfo{volume}{102}},
  \bibinfo{eid}{143002} (\bibinfo{year}{2009}).

\bibitem[{\citenamefont{Longuet-Higgins}(1961)}]{longuet61}
\bibinfo{author}{\bibfnamefont{H.~C.} \bibnamefont{Longuet-Higgins}},
  \bibinfo{journal}{Adv. Spectrosc.} \textbf{\bibinfo{volume}{II}},
  \bibinfo{pages}{429} (\bibinfo{year}{1961}).

\bibitem[{\citenamefont{Bunker and Jensen}(1998)}]{bunker98}
\bibinfo{author}{\bibfnamefont{P.~R.} \bibnamefont{Bunker}} \bibnamefont{and}
  \bibinfo{author}{\bibfnamefont{P.}~\bibnamefont{Jensen}},
  \emph{\bibinfo{title}{{Molecular Symmetry and Spectroscopy}}}
  (\bibinfo{publisher}{NRC Research Press}, \bibinfo{year}{1998}).

\bibitem[{\citenamefont{Kokoouline and Greene}(2003)}]{kokoouline03b}
\bibinfo{author}{\bibfnamefont{V.}~\bibnamefont{Kokoouline}} \bibnamefont{and}
  \bibinfo{author}{\bibfnamefont{C.~H.} \bibnamefont{Greene}},
  \bibinfo{journal}{Phys. Rev. A} \textbf{\bibinfo{volume}{68}},
  \bibinfo{pages}{012703} (\bibinfo{year}{2003}).

\bibitem[{\citenamefont{Kokoouline and Greene}(2005)}]{kokoouline05}
\bibinfo{author}{\bibfnamefont{V.}~\bibnamefont{Kokoouline}} \bibnamefont{and}
  \bibinfo{author}{\bibfnamefont{C.~H.} \bibnamefont{Greene}},
  \bibinfo{journal}{Phys. Rev. A} \textbf{\bibinfo{volume}{72}},
  \bibinfo{pages}{022712} (\bibinfo{year}{2005}).

\bibitem[{\citenamefont{{de Boor}}(2001)}]{deboor01}
\bibinfo{author}{\bibfnamefont{C.}~\bibnamefont{{de Boor}}},
  \emph{\bibinfo{title}{{A Practical Guide to Splines}}}
  (\bibinfo{publisher}{Springer}, \bibinfo{address}{New York},
  \bibinfo{year}{2001}).

\end{thebibliography}

\end{document}